# Towards an Emotion-Aware Metaverse: A Human-Centric Shipboard Fire Drill Simulator

Musaab H. Hamed-Ahmed[a], Diego Ramil-López[a], Paula Fraga-Lamas[a,b,c], Tiago M. Fernández-Caramés[a,b]

[a]*Department of Computer Engineering, Faculty of Computer Science, Universidade da Coruña, A Coruña, 15071, Spain*
[b]*Centro de Investigación CITIC, Universidade da Coruña, A Coruña, 15071, Spain*
[c]*Corresponding author: Paula Fraga-Lamas, paula.fraga@udc.es*

**Abstract**

Traditional XR and Metaverse applications prioritize user experience (UX) for adoption and success but often overlook a crucial aspect of user interaction: emotions. This article addresses this gap by presenting an emotion-aware Metaverse application: a Virtual Reality (VR) fire drill simulator designed to prepare crews for shipboard emergencies. The simulator detects emotions in real time, assessing trainees' responses under stress to improve learning outcomes. Its architecture incorporates eye-tracking and facial expression analysis via Meta Quest Pro headsets. The system features four levels whose difficulty is increased progressively to evaluate user decision-making and emotional resilience. The system was evaluated in two experimental phases. The first phase identified challenges, such as navigation issues and lack of visual guidance. These insights led to an improved second version with a better user interface, visual cues and a real-time task tracker. Performance metrics like completion times, task efficiency and emotional responses were analyzed. The obtained results show that trainees with prior VR or gaming experience navigated the scenarios more efficiently. Moreover, the addition of task-tracking visuals and navigation guidance significantly improved user performance, reducing task completion times between 14.18% and 32.72%. Emotional responses were captured, revealing that some participants were engaged, while others acted indifferently, indicating the need for more immersive elements. Overall, this article provides useful guidelines for creating the next generation of emotion-aware Metaverse applications.

*Keywords:* Metaverse, Industrial Metaverse, Emotions, Extended Reality,

Fire Drill, Maritime safety.

## 1. Introduction

The Metaverse has emerged as a transformative concept, integrating Extended Reality (XR) technologies to create immersive and interactive environments [1]. Although the Metaverse was initially conceptualized in science fiction [2], it is now being adopted across industries, including education, healthcare or entertainment [3]. One of its most promising applications is related to the field of training, specially in safety and emergency preparedness, where immersive simulations can provide effective and scalable training solutions [4, 5].

Specifically, maritime safety is a critical domain where fire emergencies pose significant risks to crew members, cargo and vessel integrity. According to the International Convention for the Safety of Life at Sea (SOLAS) regulations, regular fire drills are mandatory to ensure that crews are adequately trained to respond to onboard fires [6, 7]. However, traditional fire drills are often logistically challenging, resource-intensive and difficult to personalize for individual crew members. XR-based fire drill simulations offer an alternative by providing realistic, cost-effective and repeatable training in a controlled environment.

A crucial aspect of emergency response is emotional resilience, which indicates the ability to remain calm and to make informed decisions under pressure. However, traditional XR and Metaverse-oriented applications, although they can analyze the user inputs in terms of performance, they do not usually consider users' emotions. Thus, emotion-aware XR training enhances traditional simulations by integrating emotion-detection technologies, such as eye tracking and facial expression analysis, to assess and to adapt training scenarios in real time. By monitoring stress levels, cognitive load and decision-making behaviors, the system can personalize training to improve both technical proficiency and psychological preparedness.

This article presents the design and evaluation of an emotion-aware Virtual Reality (VR) fire drill simulator for shipboard emergencies, whose main contributions are the following:

- A holistic review is provided on the integration of XR technologies, emotion recognition methods and performance analytics to create emotion-aware Metaverse applications.

- It is described what to the knowledge of the authors is the first emotion-aware maritime safety training Metaverse application.

- A detailed evaluation is presented with groups of real final users (naval engineers) that assessed the developed solution in terms of performance and emotional engagement.

The rest of this article is structured as follows. Section 2 reviews the state of the art and the background work in relation to the Metaverse, fire drill regulations and the development of emotion-based XR solutions, as well as the most relevant hardware and software for detecting and monitoring emotions in Metaverse applications. Section 3 describes the design of the proposed emotion-aware fire drill simulator, whose implementation is detailed in Section 4. Section 5 is dedicated to the performed experiments. Finally, Section 6 details the main key findings and Section 7 is devoted to the conclusions.

## 2. State of the art

*2.1. About the Metaverse*

The concept of "Metaverse" was initially introduced by Neal Stephenson in his 1992 science fiction novel "Snow Crash" [8]. In this work, Stephenson envisioned a continuous virtual realm seamlessly integrated into everyday life. The term "Metaverse" derives from the Greek prefix "meta", meaning "beyond", and the root "verse", from "universe", thus defining the Metaverse as a reality that transcends the physical universe.

Although "Snow Crash" provides a clear description of the Metaverse, no standardized definition has been yet established [9, 10, 11]. This ambiguity allows industry leaders to shape the concept in alignment with their perspectives and strategic objectives [12, 13]. Nevertheless, despite the lack of consensus, the recognition of the Metaverse's potential across industries underscores its broad and diverse opportunities [14].

Debates surrounding the Metaverse extend to its foundational technologies [15]. For instance, the role of Augmented Reality (AR) as a core component has been debated [16]. Moreover, some authors interpret the Metaverse as a decentralized evolution of the current Internet, emphasizing user control over systems, data and virtual assets [17].

The distinction between the Metaverse and participatory Extended Reality (XR) environments further complicates its definition. Technologies such as Internet of Things (IoT) based home automation systems, which facilitate XR interactions [18], exhibit characteristics similar to those of the Metaverse but that are typically classified as Cyber-Physical Systems (CPSs) [19] rather than direct Metaverse innovations.

Despite the definitional challenges, the transformative potential of the Metaverse remains evident. It holds the promise for enabling remote collaboration and enhancing the productivity of future digital workforces, which transform into what is called Meta-Operators [20]. Through advancements in ubiquitous computing and tracking technologies, the Metaverse envisions a future in which digital and physical domains, as well as professional and commercial spheres, are seamlessly interconnected. In this context, the Metaverse represents an interactive virtual extension of the Internet, transcending traditional physical boundaries.

From a technical perspective, the Metaverse can be described as a massively scaled interoperable network of real-time rendered 3D virtual environments [21]. Users engage with the Metaverse synchronously and persistently, experiencing a continuous sense of presence while maintaining data continuity across identity, history, entitlements, objects, communications and transactions. The concept transcends any single entity or industry, with various Metaverses tailored to interests such as sports, entertainment, art or commerce [22].

In practice, multiple Metaverses coexist [23], forming collections of interconnected virtual worlds governed by hierarchical structures analogous to the Internet complex architecture and the geopolitical organization of the physical world. The terminology for these collections is still under discussion [24], with proposals including "Multiverse" [9] and "Metagalaxy" [20].

### 2.2. Fire drill regulations

Fire drills are essential for ship management and are overseen by the ship master [25]. Due to the severe threat of fire accidents to lives, cargo and financial interests, regular fire drills are mandated [26].

The International Convention for the Safety of Life at Sea (SOLAS), established in 1974 and subsequently amended, stands as the cornerstone of maritime safety and environmental protection regulations. SOLAS encompasses eleven chapters, each addressing specific aspects of maritime safety and operation [26]. Specific to fire safety, SOLAS mandates various regulations addressing fire detection, extinguishing, Personal Protective Equip-ment (PPE), evacuation procedures, and dangerous materials handling [27]. Key regulations include:

1. The availability of visual and audible alarms (Regulation II-2/7 and III 6.4.2).

2. Availability of proper evacuation through escape routes, and muster areas (Regulation II-2/12).

3. Response to fire emergency by specifying the necessary firefighting equipment, providing crew training in firefighting methods, and establishing protocols for containing and extinguishing fires (Regulation II/2 2.1.7, 5, 7.5.1, 15.2.1.1, 15.2.3, 16.2, 18.8 and III 35).

4. Regular fire drills so the crew acquaints themselves with the evacuation procedures (IMO 2009) (Regulation II-2/15.2.2 and III/19.3).

5. Availability of proper signage and navigation lighting on the escape routes (Regulation II/13.3.2.5).

Specifically, SOLAS regulation III/19.3 indicates that fire drills must be conducted monthly for each crew member, simulating realistic fire scenarios, with a focus on vulnerable ship areas [27]. Standardized fire drill scenarios typically involve several steps [28], which are illustrated in Figure 1.

A fire drill unfolds in a structured sequence, beginning with the sounding of fire alarms upon the initial alert of a fire incident. Crew members swiftly assemble at predetermined muster stations for attendance verification. Then, designated firefighters, guided by the master's instructions, prepare to confront the fire. Upon authorization from the master, firefighting personnel engage in suppression activities while prioritizing casualty rescue operations. In parallel, reserve personnel stand ready to support firefighting endeavors. Should initial firefighting attempts prove inadequate, the activation of the fixed fire extinguishing system becomes imperative, serving as a crucial backup measure in containing the escalating situation [28].

There are evaluation criteria for each step of the fire drill, assessing factors like time, performance and task completion [25]. Following drills, the ship master evaluates crew performance, providing corrective actions as needed to enhance preparedness [25].

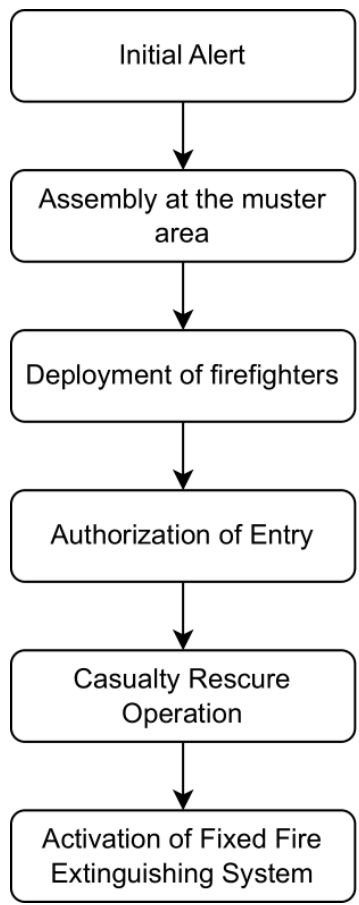

Figure 1: Steps performed for fire drills onboard ships.

*2.2.1. Metaverse Applications for Maritime Industries*

While XR technology has demonstrated effectiveness across industries, its integration into maritime operations remains limited [29]. For instance, previous research has proposed three primary uses for the use of XR in maritime operations [30]: accident prevention, accident response and accident analysis. Accident prevention involves risk assessment and management during routine operations and emergencies. Accident response focuses on evacuation processes during emergencies, while accident analysis translates documented accidents into virtual environments for analysis.

AR finds extensive use in maritime construction, aiding workshop workers during outfitting [31, 32] and assembly processes [33]. AR applications can integrate real-time sensor measurements with technology platforms like Unity, thus facilitating streamlined information retrieval processes [34, 35].

Regarding VR, it is commonly employed in training simulators due to its cost-effectiveness and environmental friendliness [36]. Applications include spray painting training [37], firefighting simulations [38], safety training [39] or dynamic risk assessment in mooring operations [40]. Multi-role [41] and multi-player [42] simulations have also been developed for lifesaving and evacuation strategies aboard passenger ships. In addition, it is worth noting that prominent maritime organizations have adopted VR technology for training and educational initiatives, spanning safety protocols, inspection procedures and onboard operations [39].

*2.3. Emotion detection for XR*

*2.3.1. Detecting emotions with XR technologies*

Emotion detection can become a fundamental component for XR applications where user engagement, stress response and decision making are critical. In immersive training environments, such as a Metaverse-based fire drill simulator, monitoring emotional states provides valuable insights into user behavior under pressure. By integrating emotion recognition mechanisms, these systems can dynamically adjust training scenarios and enhance realism, learning outcomes and the overall preparedness for real-world emergencies.

The detection of emotional states in immersive environments relies on physiological and behavioral data analysis, encompassing facial expressions, eye movement tracking, heart rate variability and galvanic skin response [43]. These bio signals offer real-time feedback on user engagement and affective states, enabling systems to deliver more adaptive and responsive experiences.

However, emotion recognition in virtual settings presents several challenges, including inter-individual and cross-cultural variations in emotional expression and sensor accuracy limitations [44]. Differences in physiological and psychological factors influence the manifestation and interpretation of emotions, necessitating the use of diverse datasets to enhance system reliability. Moreover, the precision of emotion detection depends on hardware quality, including cameras, eye tracking systems and wearable sensors, which may introduce constraints in dynamic environments.

*2.3.2. XR emotion detection techniques*

Emotion recognition in immersive environments is achieved through multiple techniques, including:

- Facial expression analysis: tracking facial movements enables the inference of emotional states based on expressions such as frowning, smiling or eyebrow motion [45]. Currently, only a few head-mounted displays incorporate inward-facing sensors to capture subtle facial cues, improving avatar realism and scenario adaptation [46].

- Eye tracking: the analysis of gaze patterns, pupil dilation and blink rates provides insights into stress levels, cognitive load and emotional engagement [44]. These data are particularly valuable for assessing decision making processes in high pressure simulations.

- Physiological signal monitoring: wearable sensors are able to measure physiological indicators such as heart rate, skin conductivity or temperature fluctuations to detect emotional arousal [47]. These metrics help when evaluating stress and engagement levels in training scenarios.

- Voice analysis: variations in speech tone, pitch and rhythm can reveal emotional states such as urgency, hesitation or confidence, making this method particularly useful for assessing verbal responses in interactive training exercises [44].

*2.4. Emotion-aware applications on the Metaverse*

The Metaverse offers substantial potential for the development of emotion-driven applications, which enhance user interaction and engagement within virtual environments. By integrating emotional responsiveness into these digital spaces, platforms can create more immersive and adaptive experiences that respond to users' emotional states. Such applications are proving to be beneficial across a range of sectors, including industrial training, healthcare and collaborative workspaces, where they improve the realism of simulations, support safety protocols, and enable a higher degree of personalization.

For instance, in the healthcare domain, platforms like XRHealth have incorporated emotion-sensitive avatars to support therapeutic interventions [48]. An example of use case is the treatment of post-traumatic stress disorder, for which XRHealth adjusts exposure therapy based on the user's emotional feedback, employing facial expression analysis to modify the intensity of the session. This personalized approach fosters a controlled yet dynamic environment that adapts to the patient's emotional needs, facilitating recovery in ways that traditional therapy may not be able to achieve. In addition, virtual environments can be utilized for rehabilitation, where patients recovering from injuries or surgeries interact with scenarios that are adjusted according to their emotional responses, providing both physical and emotional support.

Moreover, it is worth indicating that Artificial Intelligence (AI) driven virtual environments are revolutionizing mental well-being interventions. For example, the Merrytopia project [49] combines natural language processing and Machine Learning to analyze users' journal entries and emotional states. Such a platform generates personalized recommendations and immersive mindfulness scenes, dynamically adjusting interactions through real-time sentiment analysis. This creates a positive feedback loop where an AI-driven chatbot reinforces constructive aspects of users' experiences, offering scalable solutions for stress and anxiety management. Thus, the developed system demonstrates how adaptive virtual environments can complement traditional therapeutic methods by tailoring support to individual emotional needs.

Beyond healthcare, the Metaverse is also being leveraged to enhance emotional intelligence and interpersonal skills. An example is a study that introduced a Metaverse-based English teaching solution that makes use of emotion-based analysis methods to personalize the learning experience [50]. This approach creates an immersive and interactive teaching environment, enabling educators to adjust lessons based on students' emotional states, thereby increasing engagement and improving learning outcomes.

In the educational field, XR-based systems can harness multimodal data to optimize learning processes. For instance, previous research has made use of eye-tracking, facial expression analysis and electrodermal activity sensors to demonstrate that reduced blink rates and sustained visual attention on task-relevant content correlate strongly with improved knowledge retention [51]. These findings enable the design of adaptive systems that respond to cognitive load indicators in real time, thus enabling to adjust the complexity of procedural training scenarios.

## 2.5. Hardware and software for detecting emotions in Metaverse Applications

### 2.5.1. Hardware

Effective emotion detection in industrial Metaverse environments relies on advanced hardware and intelligent software systems. Modern AR and Mixed Reality (MR) headsets, such as Meta Quest Pro [52] or Microsoft HoloLens [53], integrate high resolution cameras (up to 13 MP) and sophisticated eye tracking technologies (e.g., Vive Focus 3 Eye Tracker [54]). Such headsets are able to monitor gaze direction and pupil dilation, which are key indicators of cognitive states like focus and fatigue. In addition, facial tracking devices, such as the HTC Vive Facial Tracker [55], enable real-time detection of microexpressions and lip movements, helping to infer emotions like happiness, surprise or stress.

Specialized accessories further enhance emotion-recognition capabilities. Haptic gloves, such as Teslaglove [56], incorporate physiological sensors that measure heart rate and blood oxygen levels, providing insight into stress and anxiety. Full body tracking devices, including Sony Mocopi [57] and HaritoraX [58], analyze posture and movement patterns, which can be linked to emotional states (e.g., a slouched posture indicating frustration). Omnidirectional treadmills like Virtuix Omni One [59] track locomotion behaviors, correlating movement speed and patterns with agitation or calmness.

### 2.5.2. Software: emotion-aware applications

Emotion-aware applications can leverage real-time emotional data to dynamically adjust system behavior, enhancing user experience, safety and efficiency. Such applications integrate biometric and affective computing technologies to tailor interactions based on the emotional states of the users. By using advanced haptic feedback systems, such as bHaptics TactSuit X16 [60], immediate tactile responses can be provided, including chest vibrations during high-stress situations, which can help in user guidance and decision making. Moreover, emotion adaptive visual interfaces enhance situational awareness by adjusting lighting and color schemes in immersive environments to maintain focus and to reduce cognitive overload.

Emotion-aware applications have broad applicability across various industrial domains. For instance, wearable biometric devices, such as Teslaglove, can monitor the stress and fatigue levels of the users, thus triggering automated safety protocols, including break notifications or task reallocation, to prevent workplace accidents. These devices also contribute to long-term wellness monitoring, enabling predictive analytics to anticipate potential health risks before they manifest as critical issues.

## 3. Design of an Emotion-aware Metaverse Application: a Fire Drill Simulator

### 3.1. Objective of the application

A fire drill simulator helps to carry out training for fire emergencies that might occur on board vessels. Its objective is to enhance trainees' decision-making and risk-assessment capabilities by simulating various fire scenarios and allowing rehearsal of key emergency procedures in a controlled manner.

Adding the component of emotional control within simulated procedures enhances realism, considering real fire emergency situations certainly would induce emotions like fear and stress, which could affect performance. With the incorporation of emotional recognition, the system not only assesses procedural competencies but also emotional resilience, thus providing personalized feedback and enhancing preparedness for high-pressure situations.

### 3.2. System architecture

Figure 2 depicts the communications architecture of the proposed system. Such an architecture represents the traditional structure of a Metaverse application: the users (i.e., Meta-Operators) wear Metaverse-ready XR devices that act as gateways to enter the different Metaverses provided by the available MetaGalaxies. Some of such MetaGalaxies are included as an example, but there are many more, which can be tailored to specific fields (e.g., industry, entertainment, gaming).

### 3.3. Design of the emergency scenario

Since crafting scenarios for every conceivable ship area proves impractical, previous research [61, 62] has underscored the heightened likelihood of fire occurrences in specific ship zones, notably the engine room and the accommodation area, particularly the galley. These areas are prone to fire incidents due to factors such as the presence of machinery and electrical equipment in the engine room, along with the abundance of oil and fuel, which pose ignition risks. Similarly, electrical faults in the galley can swiftly escalate into fire outbreaks. Consequently, the engine room and galley were selected as focal points for scenario development, aligning with empirical evidence of fire prevalence in these ship sections.

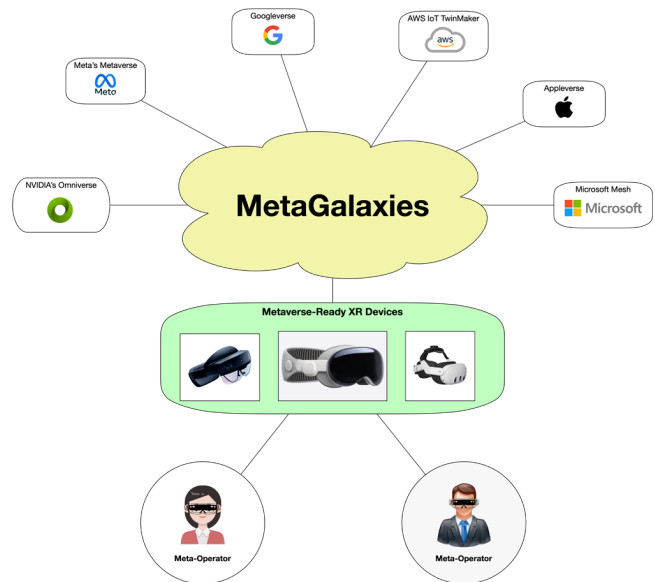

Figure 2: General communications architecture.

When crafting the scenario, it was considered that the main objective was to create an immersive training experience for crew members to simulate fire emergencies and to evaluate their response capabilities, behavior and risk analysis skills. Insights gathered on shipboard fire drill scenarios [25, 28] inform the design process, supplemented by guidance from a fire drill training and operation booklet [63]. The scenario, developed with the previously described critical elements in mind, unfolds as follows:

1. The Meta-Operators to be trained navigate freely within the virtual environment, situated either in the engine room or the galley.

2. Their task involves promptly identifying the fire's location, signaled by auditory cues like burning sounds or visual cues such as flames.

3. Upon detection, participants inform the ship master about the fire location and size, and activate the fire alarm system.

4. They then assess the fire's severity to determine whether it is controllable and extinguishable, or whether it poses an imminent threat.

5. Finally, participants evacuate to the designated muster area.

Following such stages, the simulator takes the trainees within different levels where the intensity of the scenario increases. This concept draws inspiration from the notion of gamification [64, 65, 66], which underscores the effectiveness of games in enhancing problem-solving skills and facilitating learning. In the implementation presented in this article, gamification is included through the different stages within the application, each presenting diverse fire intensity levels and fire eruption locations. Thus, the proposed multi-level approach serves as an effective means of evaluating the trainees' performance during the whole training process.

## 4. Implementation of the Metaverse application

### 4.1. Hardware and software

Among the potential headsets to be used, Meta Quest Pro was selected thanks to its ability to monitor natural facial expressions and to track the user eyes. Specifically, the Quest Pro features advanced eye-tracking technologies use inward-facing sensors to monitor and to interpret eye movements in real time.

Regarding the selected software, Unity Editor version 2022.3.15f1 was used with the Universal Rendering Pipeline (URP). The XR Interaction Toolkit facilitated the creation of immersive interactions within the application. The OpenXR Software Development Kit (SDK) was employed for communicating the Meta Quest Pro with the developed application, ensuring cross-platform compatibility and future-proofing [67]. The Face Tracking Application Programming Interface (API) from the Meta Movement SDK was used, since it allows developers to map detected expressions onto these characters for natural interactions. Thus, such an API converts movements into expressions based on the Facial Action Coding System (FACS).

In addition, Blender was used alongside Unity for specific segments of the virtual environment due to its robust feature set and VR readiness [68].

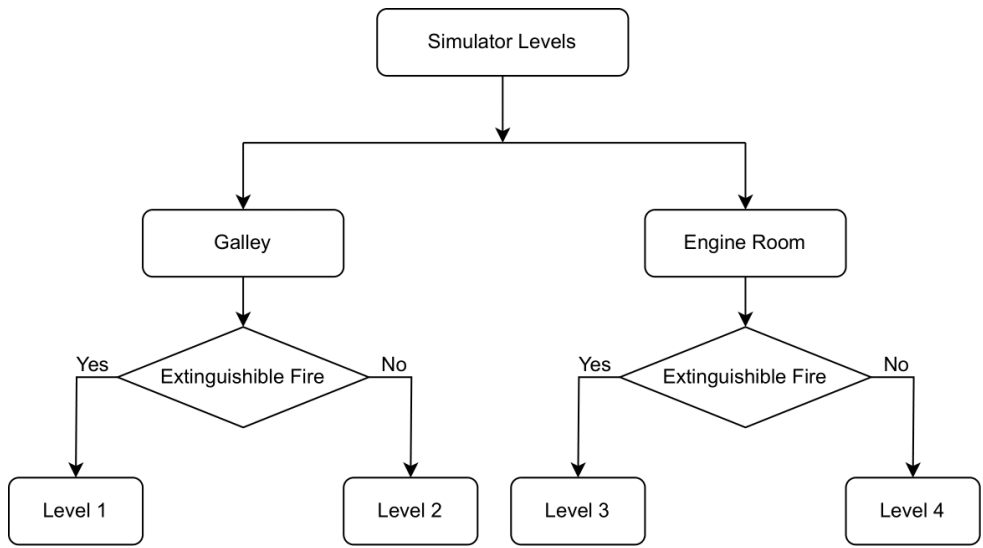

Figure 3: Levels of the fire-drill simulator.

*4.2. Scenario levels*

To construct the devised scenario, four main scenes were identified, with each scene representing a level in the simulator as shown in Figure 3.

Initially, in the first level, trainees are introduced to a scenario where a small fire erupts in the galley area, offering a manageable firefighting situation to acquaint them with basic fire management techniques. In this level, guidance text is provided to the trainee to perform the required tasks (as it is illustrated in Figure 4). As they advance to the second level, the intensity of the fire escalates, presenting a heightened challenge with an inextinguishable blaze, thereby testing the trainees' adaptability and decision-making under pressure. Moving on to the third level, a new challenge emerges as a fire breaks out in the engine room, requiring trainees to navigate through the scenario without explicit instructions, relying instead on established safety protocols accessible via the User Interface (UI) instruction menu. Despite the increased complexity, the fire remains extinguishable at this stage, fostering critical thinking and procedural adherence. Finally, the fourth level presents the maximum difficulty, featuring a substantial fire outbreak in the engine room that rapidly intensifies. Here, trainees face an inextinguishable inferno, presenting the ultimate test of their firefighting prowess and resilience in the face of escalating adversity.

To assess the performance of the trainees, a script was created to log the critical interactions of the trainees with the required game objects. Thus, every interaction is recorded together with a timestamp and with the event, whether it was grabbing an object or activating it. This record later allows for estimating the user workload during the training.

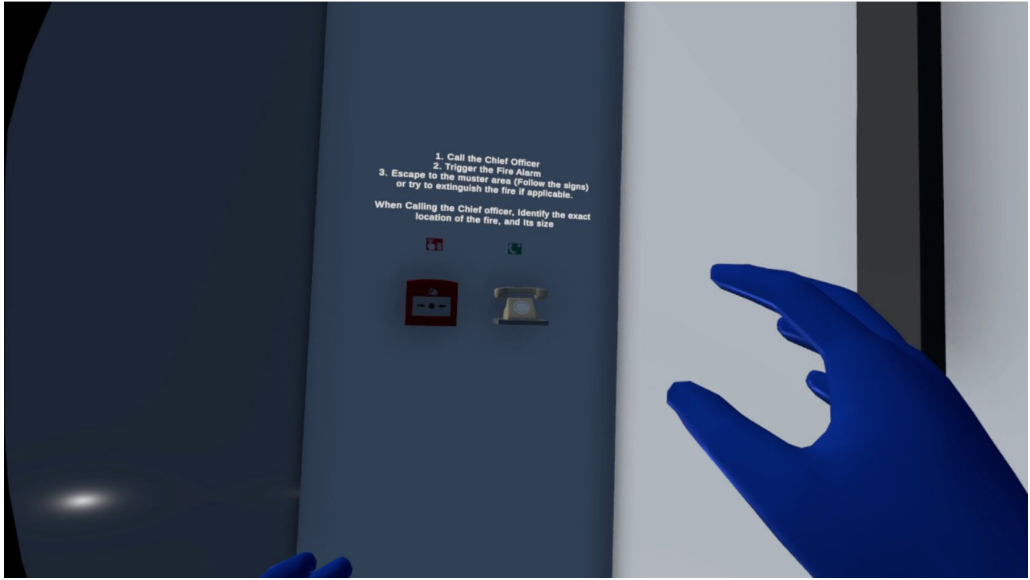

Figure 4: Example of guidance text in Level 1.

### 4.3. Virtual environment

The implementation of the virtual scenario involved importing a cargo ship model sourced and constructing the ship interior cabin. Blender, in conjunction with the HomeBuilder add-on, facilitated the design, inspired by a tween-decker general cargo ship arrangement [69]. For instance, Figures 5 and 6 show, respectively, the created galley area and spawn room. Moreover, Figure 7 shows the engine room.

Lighting was applied using Unity's point light game object for optimal visibility. Collider components were implemented to enhance interaction realism. Interactive elements such as fire extinguishers, alarms and emergency phones were included. Simulated fire and smoke effects were integrated by using Unity particle system. As an example, Figure 8 shows the implemented fire extinguisher when it was used by a trainee to put off a simulated fire.

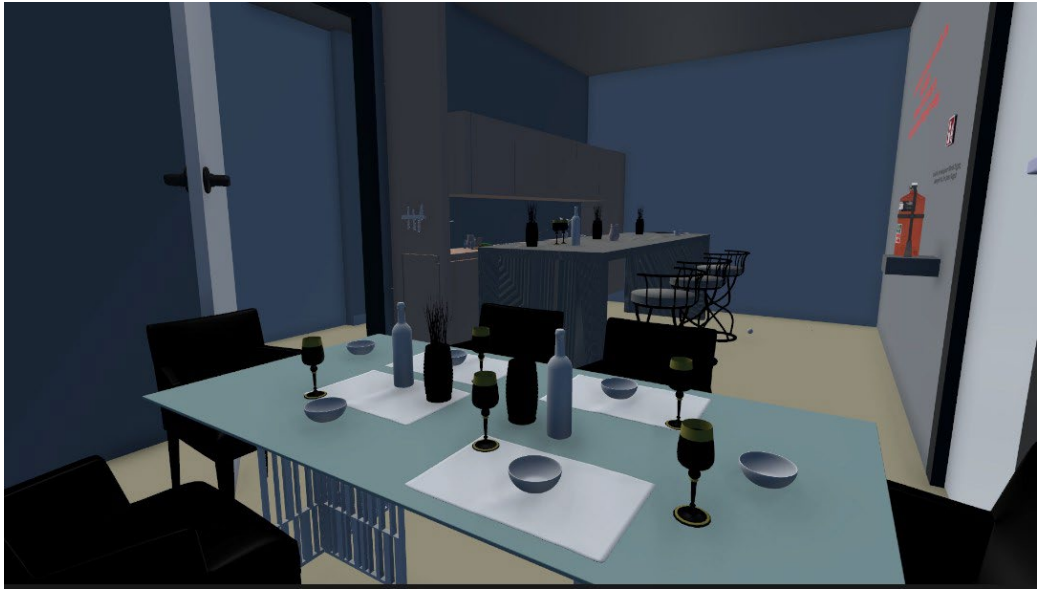

Figure 5: Galley area of the ship.

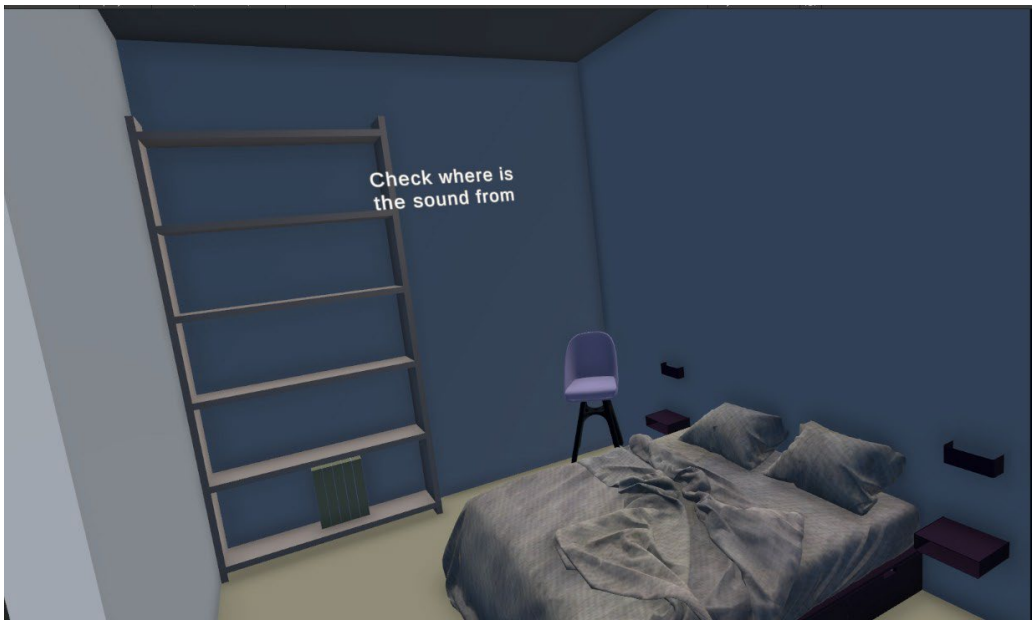

Figure 6: Spawn room of the application.

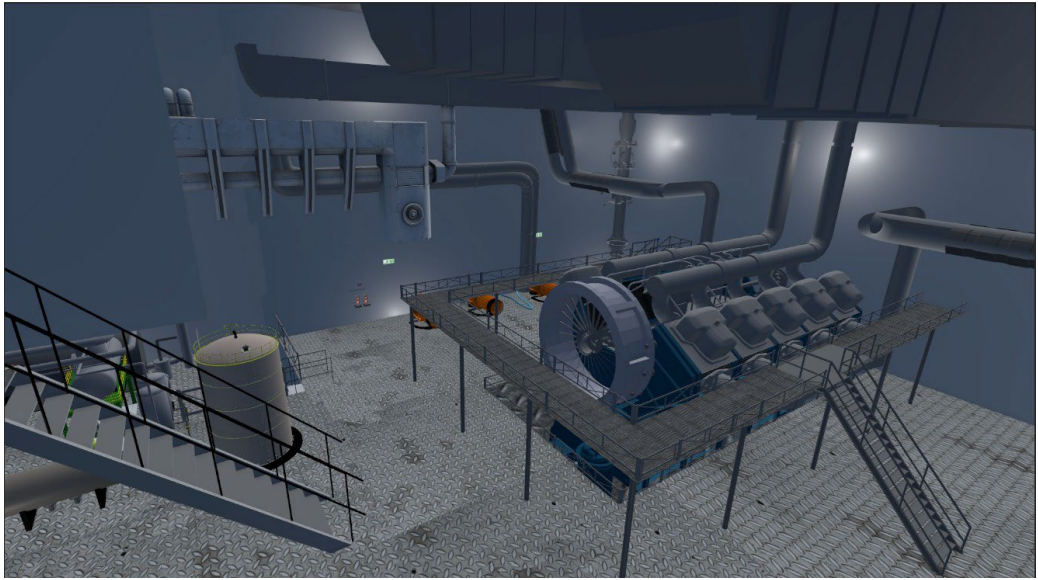

Figure 7: Engine room of the ship.

Finally, a Sky High Dynamic Range Imaging (HDRI) scene and an ocean shader from Unity were included to add realism to the scenario (these can be observed in Figure 9).

*4.4. Emotion detection*

Emotion detection focused on two separate features: eye tracking and natural facial expressions. To assess the workload of the trainees, the following actions were performed:

- An eye tracker script was implemented using Meta Movement SDK, with its built-in Eye Gaze script. Moreover, another script was created to ensure the application of the appropriate raycast and proper naming of the game objects.

- Facial expressions were logged with the Meta Movement SDK and the built-in Face Expressions API. Such an API translated the camera and the inward-sensors into blendshapes that monitored the logged expression with its strength thorough the respective weighting of each expression. In addition, a second script to process the weighted expressions to determine through a rule-based logic the detected emotions (e.g., happiness, sadness, surprise).

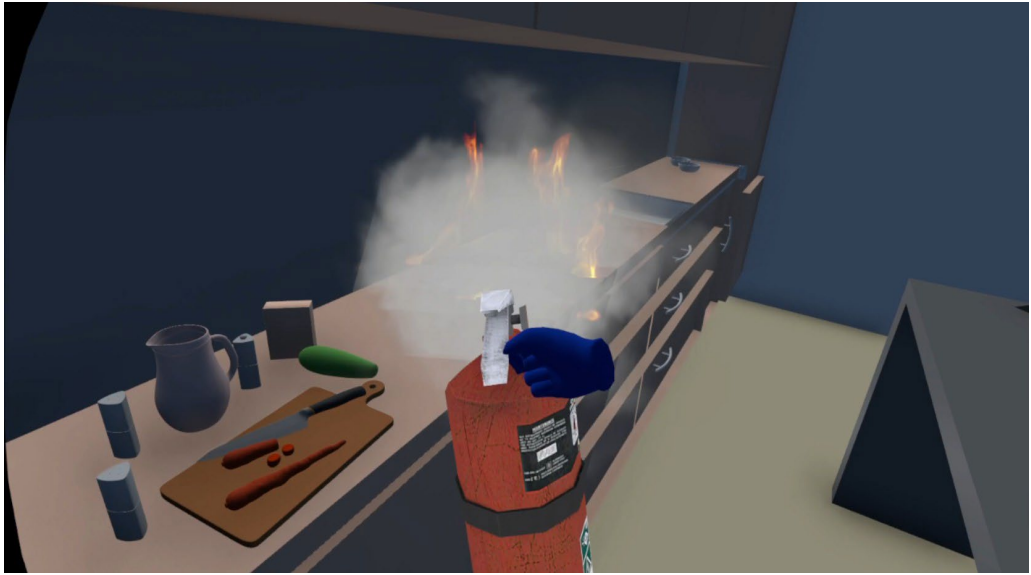

Figure 8: Moment when a trainee performed the fire extinguishing process.

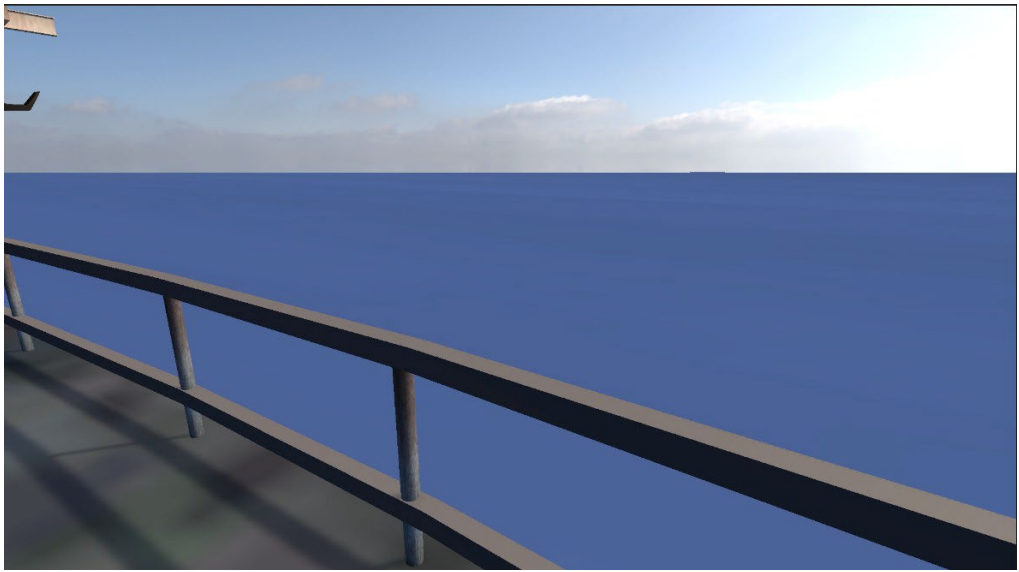

Figure 9: Ocean shader and sky scene added to the simulator.

- Lastly, the readings from both the eye tracking and the face expressions were stored simultaneously in a dataset for each level of the simulator. In this way, the logging mechanism allows for determining when the trainee is gazing at an object, and associates such an action with the detected emotions.

## 5. Experiments

*5.1. Description of the experiments*

Two sets of experiments were conducted in order to evaluate two different versions of the application:

- The first experiment was conducted with an earlier version of the application that only included textual cues to help the trainees through the tasks, but no visual cues were provided on how to perform the tasks (neither a map nor navigation clues to explore the virtual map). The first version measured the completion time to finish each level. Ten naval engineers (8 men, 2 women) from the Erasmus Mundus Joint Masters Degree Sustainable Ship and Shipping 4.0 (SEAS 4.0) (imparted by the University of Napoli Federico II, the University of A Coruña and the University of Zagreb), participated in the first experiment. Before the tests began, each participant was instructed on how to use the controllers, the purpose of the simulator, and on how to conduct the fire drill.

- The second experiment was conducted on a modified version of the simulator. The new version was based on the outcomes and on the feedback collected during the first experiment, which focused on (1) shortening the time of the fire extinguishing process on the first level, (2) reducing the navigation speed to slightly mitigate motion sickness, (3) introducing visual navigation guidance to help the trainees navigate through the virtual environments, and (4) providing visual feedback cues when the tasks were completed. In addition, a comprehensive task logger was added to the application to gain more insights on the behavior and workload of the trainees during the use of the simulator, as well as introducing emotion detection to complement the analysis of the workload. For this second experiment, 7 naval engineers (4 men, 3 women) from the SEAS 4.0 participated (3 of them tested the simulator for the first time). Like for the first experiment, the testers, were introduced to the use of the controllers and on the fire drill protocols.

For both experiments, the testers were asked to provide their knowledge level and experience with fire drills, video games and VR, since such experience can influence the obtained results. Such data is shown in Table 1 and Table 2.

| Tester No. | Gender | Previous Experience in Fire Drills | Previous Experience in VR | Previous Experience in Playing Video Games |
|---|---|---|---|---|
| 1 | Man | High | High | High |
| 2 | Man | High | Low | Low |
| 3 | Woman | Low | Low | High |
| 4 | Man | Medium | Low | Low |
| 5 | Woman | Low | Low | High |
| 6 | Man | Low | Low | High |
| 7 | Man | Low | Low | Low |
| 8 | Man | High | Low | High |
| 9 | Man | Low | Low | High |
| 10 | Man | High | Medium | High |

Table 1: Background of the testers of the first experiment.

| Tester No. | Gender | Previous Experience in Fire Drills | Previous Experience in VR | Previous Experience in Playing Video Games |
|---|---|---|---|---|
| 1 | Man | High | High | High |
| 2 | Woman | Medium | Low | High |
| 3 | Woman | High | Low | Low |
| 4 | Man | Medium | Medium | Medium |
| 5 | Man | Low | Low | Low |
| 6 | Woman | Medium | Medium | Low |
| 7 | Man | Medium | Medium | High |

Table 2: Background of the testers of the second experiment.

## 5.2. Performance Tests
### 5.2.1. Time required to complete each level

During the first experiment, as it was previously mentioned, the only collected data was the time to complete each level, which is depicted in Figure 10. As it can be observed, testers spent a lot of time on Level 1, which is essentially due to the fact that the implemented fire extinguishing process took a significant amount of time (52 seconds when performed perfectly), and since it was the first contact of the testers with the application, so they needed to adapt to the controls and to the virtual environment. A similar situation occurred in Level 3, where the testers needed to adapt to a new scenario (the engine room).

One conclusion that can be drawn from the experiments is that the testers with prior experience in video gaming were the fastest to complete the tasks and levels compared to the testers with low video gaming experience. For instance, an experienced gamer like Tester No. 1 required clearly less time than less experienced gamers. In fact, non-experienced gamers required on average roughly the double of the time of experienced gamers.

Another observation concerning task completion is the fact that all testers followed the exact sequence of instructions provided by the fire drill, except for testers 4, 8 and 9. Specifically, testers 4 and 9 attempted to extinguish the fire in Level 2, while tester 8 evacuated before extinguishing the fire in Level 3. This suggested a need for a real-time task tracker (later implemented for the second experiment) to guide trainees when they do not accomplish a specific task. Nonetheless, overall, most testers improved their decision-making skills in assessing fire severity and determining whether to extinguish the fire or to evacuate.

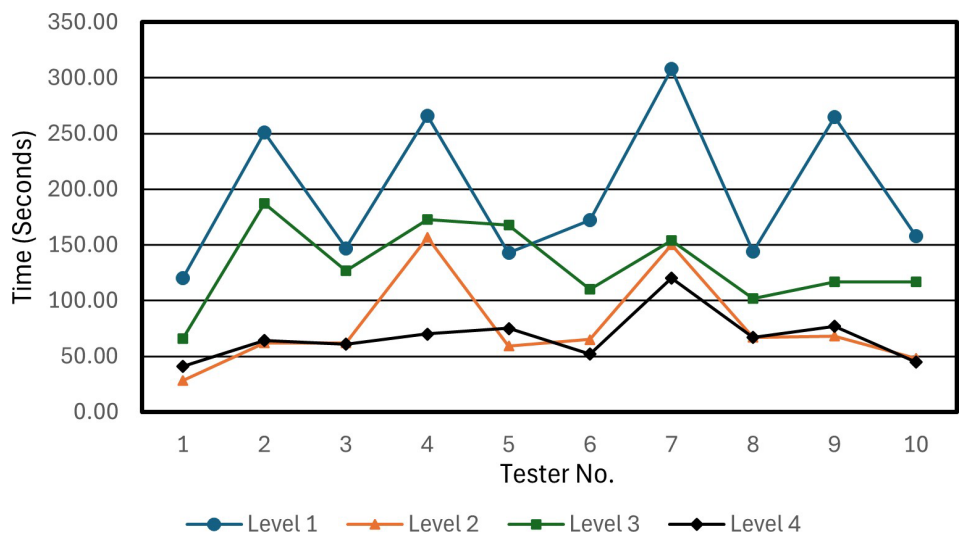

Figure 10: Time spent on each level for the first experiment.

For the second experiment, in order to accelerate the training process, the time for extinguishing the fire was drastically reduced from 52 seconds to 7 seconds. In addition, the navigation speed was adjusted and visual cues (arrows) were added to enhance the guidance. With such modifications, the obtained times to finish each level were significantly cut down, as it can be observed in Figure 11.

It is important to note that Testers 2, 4 and 5 evaluated the simulator for the first time, so they required (specially testers 4 and 5) of significantly more time that the other testers. Moreover, it is worth pointing out that Testers 5 and 6 barely had any experience in video gaming and VR in comparison to Tester 2.

On the other hand, testers who participated in the first experiments performed better in the second experiment. This is due to the previous training, but also thanks to the additional help provided by the included additional visual guidance.

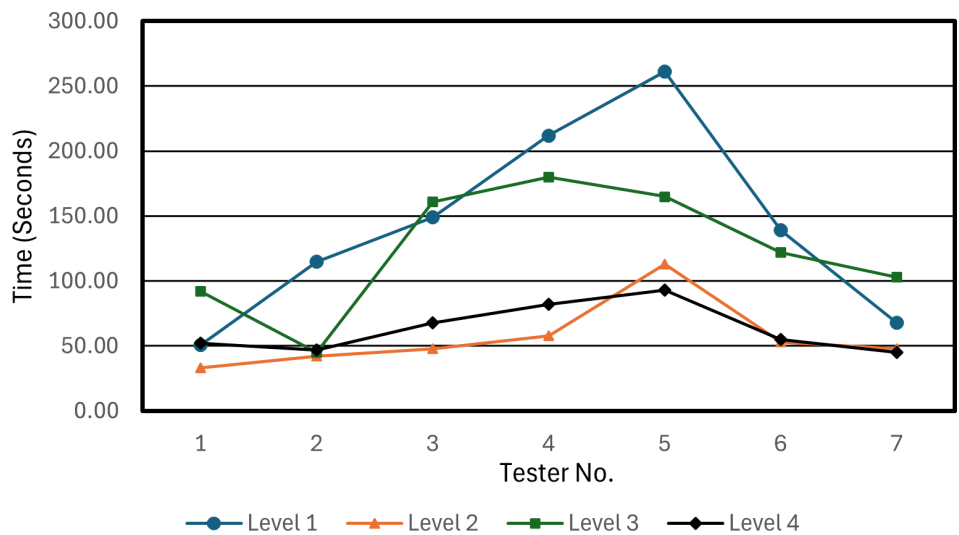

Figure 11: Time required to finish each level in the second experiment.

*5.2.2. Number gazed objects*

For this comparison, the data from the second experiment was utilized to measure the amount of objects gazed by each tester. The gazes of Tester 1 were used as a reference due to his larger experience on onboard fire drill operation and VR. Moreover, the eye gaze data were pre-processed to eliminate the effect of eye blinking.

Figure 12 shows the number of gazes per tester. It can be observed that the testers stared at a lot of objects in levels 1 and 3, since such levels require exploring virtual environments that are new to the testers. Specifically, 19.38% of the gazes were aimed at the fire collider, 17.51% at the virtual phone collider, 14.46% at the fire extinguisher, and 10.7% at the fire alarm. This is normal because the previously mentioned objects are the main objects involved in the fire drill. However, it is worth noting that only 8.33% of the gazes were aimed at watching the signage (e.g., fire exits, muster area, emergency phone signs or the fire alarm sign) and 6.12% at the guiding text. This indicated a need for providing highlighted interactive elements on the signage and on the guiding text to attract the tester's gazes and to make the simulator more intuitive.

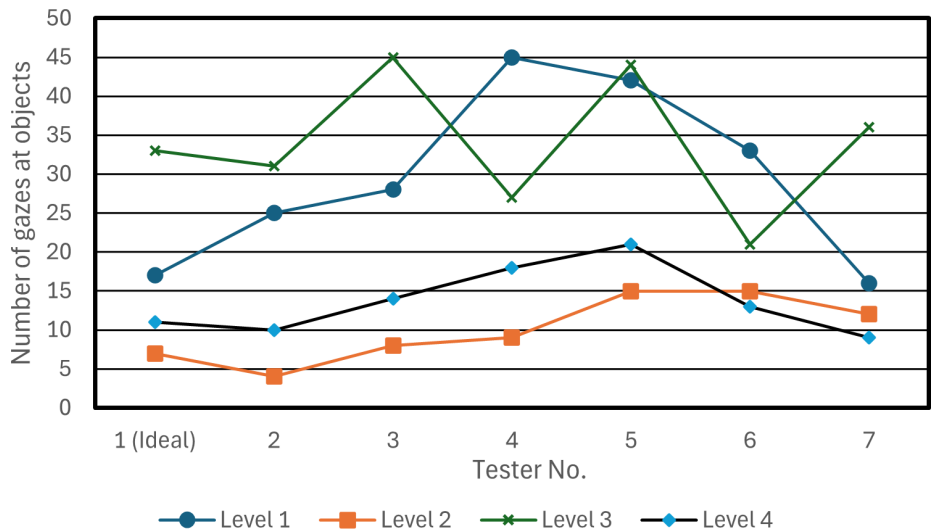

Figure 12: Number of the gazed objects for each tester and level.

### 5.2.3. Sequence of gazed objects

In order to determine the similarity of the sequences of gazed objects, an approach used in bioinformatics for comparing genome sequences was used. Specifically, the Longest Common Subsequence (LCS) and the Sliding Window (SW) techniques were implemented. The LCS is the longest common subsequence between two sets of sequences. This method utilizes dynamic computing to construct a matrix with the length of the longest common subsequences [70]. The subsequence is generated by following the sequence order and neglecting unmatching elements, and it is preferred when following the order of the sequence is necessary. Thus, the similarity score with LCS is calculated by [71]:

$$\text{Similarity}_{\text{LCS}} = \sqrt{\frac{\text{LCS Length}}{\text{Ideal Length} \times \text{Compared Length}}} \quad (1)$$

where:

- LCS Length is the length of the longest common subsequence between the sequence of gazed objects obtained by the reference tester (Tester 1 in these experiments) and the sequence of the tester to be compared.

- Ideal Length is the number of elements in the sequence of the reference tester.

- Compared Length is the number of elements of the sequence obtained by the tester to be compared.

On the other hand, SW offers a less strict approach to calculating the similarity score, where a window of a specified size is used to slide over the ideal sequence to find matches within the user sequence. This approach is effective in identifying near matches rather than requiring perfect alignment [72]. Specifically, the similarity score for SW is computed as follows:

$$\text{Similarity}_{\text{SW}} = \sqrt{\frac{\text{Match Count}}{\text{Ideal Length} \times \text{Compared Length}}} \quad (2)$$

where:

- Match Count is the length of the longest common subsequence between the sequence of the reference tester and the sequence of the tester to be compared.

- Ideal Length is the number of elements in the sequence of the reference tester.

- Compared Length is the number of elements in the sequence of the tester to be compared.

Figure 13 shows the similarity score obtained when using LCS for the users that carried out the second experiments described in Section 5.1. In such a Figure it can be observed that the sequence similarity score of the levels tends to oscillate around 0.5: level 3 has an average similarity score of 0.552 and level 2 average score is 0.448, while levels 1 and 4 both average 0.506 and 0.508, respectively. Moreover, the average for each tester for all four levels fluctuates between 0.488 and 0.56.

It must be noted that the obtained results are conditioned by the definition of LCS, which is quite rigid (i.e., it is used to determine if two sequences are exactly the same), when, in practice, during the experiments, the testers' gazes deviated from the ideal sequence due to the lack of indicators to follow a specific sequence of objects. Thus, this resulted in relatively low similarity scores when comparing with the ideal sequence.

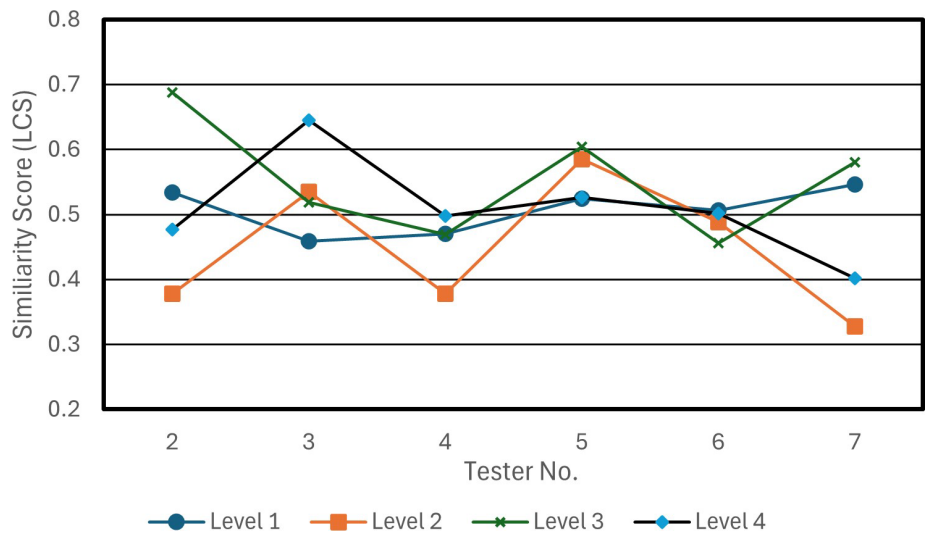

Figure 13: Similarity score for LCS.

Figure 14 shows the similarity scores obtained when using SW for a window size of 2 (such a size is necessary due to the non-colocalization of the gazed objects in the virtual environment). As it can be observed in Figure 14, the average similarity score is lower than for LCS scores. This is essentially due to the small window size of 2: it requires strict close alignment precision, which is not flexible when gaze deviations occur.

Specifically, level 1 has the lowest average similarity score (0.292), while level 3 has the highest (0.472) (such a metric is 0.349 for level 2 and 0.425 for level 4). These results mean that levels 3 and 4 (the ones centered in the engine room) are more intuitive when navigating and interacting with objects.

Testers average a SW value between 0.285 and 0.506, which allows for estimating how far they are from learning the ideal sequence of actions to be taken in the simulated fire drill.

Overall, after analyzing the LCS and SW values, it can be concluded that, in order to improve them, it is necessary to provide some sort of visual cues or guiding audio to help users during their training.

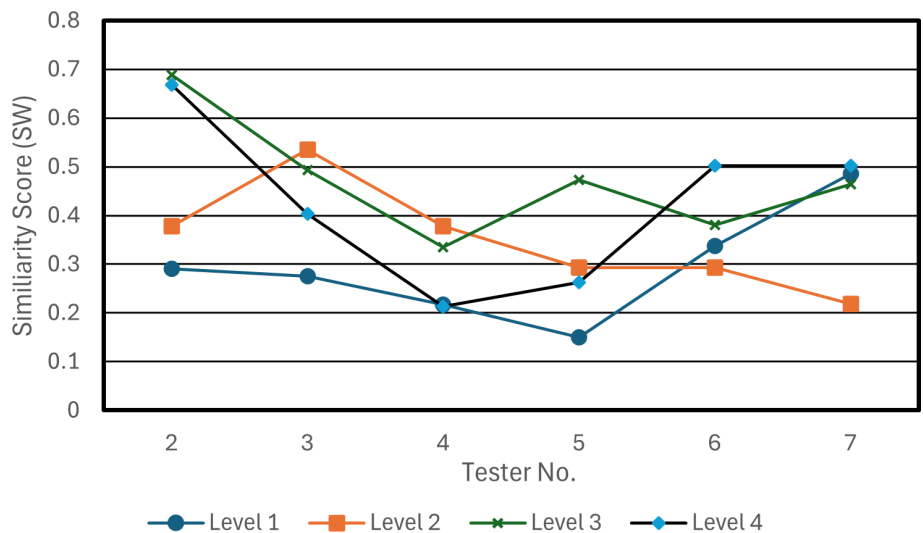

Figure 14: Similarity score for SW.

*5.2.4. Performance with and without adding visual cues*

To assess the performance improvement achieved after including additional visual navigation guidance (text and virtual arrows), the mean completion times and standard deviations for each level were compared. The results are shown in Table 3, where it can be observed that significant improvements were obtained. Specifically, on average, completion time was reduced between 14.18 % and 32.72 %, thus demonstrating that the included visual clues were clearly effective.

Moreover, 15 shows box plots that allow for illustrating the performance of the testers before ("old levels") and after including additional visual guidance ("new levels"). As it can also be observed in Table 3, Level 2 shows the most significant decrease in both mean time and standard deviation.

| Level | Old Implementation Mean (s) | Std Dev (s) | New Implementation Mean (s) | Std Dev (s) | % Improvement |
|---|---|---|---|---|---|
| 1 | 166.88 | 73.84 | 142.14 | 74.84 | **14.82%** |
| 2 | 83.88 | 46.27 | 56.43 | 26.18 | **32.72%** |
| 3 | 144.65 | 41.13 | 124.00 | 48.12 | **14.31%** |
| 4 | 73.58 | 24.24 | 63.14 | 18.49 | **14.18%** |

Table 3: Completion times and standard deviation of old and new implementations.

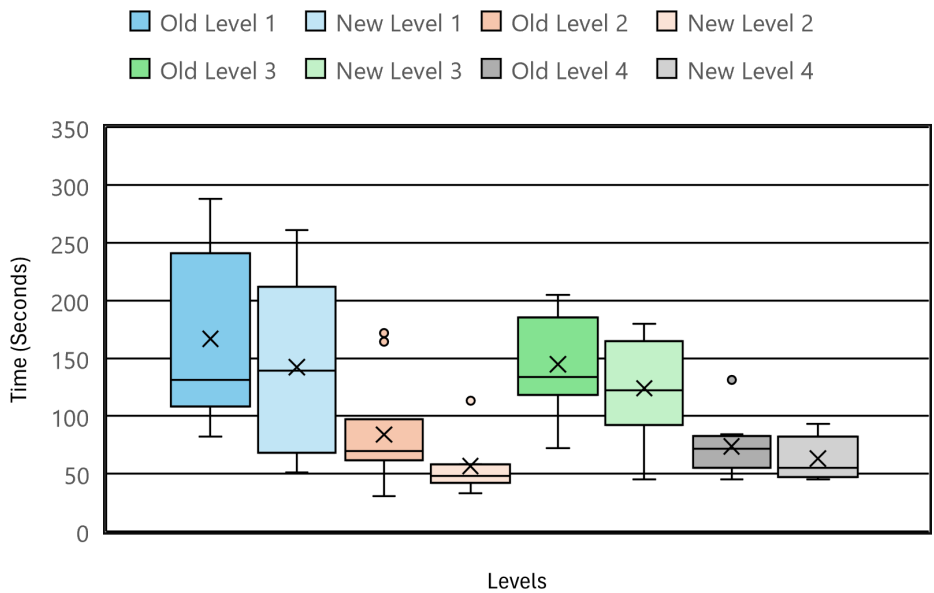

Figure 15: Box plot comparison of the performance of testers with and without the added visual cues.

## 5.3. Emotion Detection Performance

### 5.3.1. Experimental setup

In order to evaluate the emotional performance of the testers, the Facial Action Coding System (FACS) was utilized to interpret their raw expressions to detect valid emotions. The FACS is a system developed for classifying human facial expressions based on muscle movements. It was originally created by Paul Ekman and Wallace V. Friesen in 1978 [73] and is now widely used in psychology, neuroscience or artificial intelligence [74]. The framework actually identifies facial muscle movements that can be combined to infer emotions (e.g., happiness, sadness, anger, surprise). For instance, genuine smiles, which express happiness, are characterized by raising cheeks and pulling the corner of the lips, while wrinkling the nose and raising the upper lip are related to disgust.

The detected emotions were recorded in conjunction with the objects the testers were gazing at, what helped to assess with accuracy the reactions of the testers in certain situations and when using specific objects. Thus, the detected facial expressions are used to interpret the level of "good" or "bad" emotions felt by each tester during the simulation.

*5.3.2. Emotion detection accuracy*

To conduct this test a list of facial expressions was mapped to each game object, and then assessed if the tester has made an expression that fits the list criteria. For instance, when a user looked at a fire, it was expected to detect fear. Thus, the more the tester reacted with an expression that fitted the list criteria, the higher the emotion detection accuracy, which was measured by dividing the number of correctly detected expressions by the total number of detected expressions.

It is worth mentioning that, although the application was recording continuously the facial expressions of the testers, such expressions cannot always be related to an emotion as defined by the FACS. In such situations, the application recorded that the system detected "no emotion". This is important to understand the difference between Figures 16 and 17, which show the different accuracy levels (i.e., the similarity of the detected emotion with respect to the one indicated in the list criteria in relation to a specific object) collected for all the participants in the second experiment described in Section 5.1. In practice, this means that the data depicted in Figure 16 was pre-processed to remove when "no emotion" was detected when staring at an object. Note that such a detection of "no emotion" can occur during the whole interaction with the object (what would be accounted as an incorrect emotion detection if that is different from what is indicated in the pre-established list of emotions related to the object) or at some specific time instants during the interaction (for example, a tester can first express fear when using a fire extinguisher to put out a fire in the engine room, but then, as the fire is being put out, the tester may express some sort of relief that can be detected as "no emotion" by the Meta Quest Pro). In contrast, in Figure 17, such "no emotion" events are also accounted, so the emotion detection accuracy shows significantly lower results respect to Figure 16.

Considering the previously described experimental conditions, it can be observed in Figure 16 that Tester 1 shows the highest recorded emotion accuracy according to the established list criteria for all the levels when excluding the "no emotion" events. However, such a tester accuracy drops when including the "no emotion" events, which indicates that, although the tester reacted to the different situations, overall he did not express as many emotions as expected during the test. Regarding Tester 2, she expressed the expected emotions on levels 3 and 4, achieving also high accuracy when including the "no emotion" events.

With respect to the rest of the testers, although they expressed different emotions when using the application, they did not reacted as expected, so, overall, their accuracy levels in both Figures 16 and 17 is overall low. This is essentially related to how the pre-defined list of emotions was defined, which diverges from the actual reactions of the testers. For instance, when putting out a fire, it was expected to detect fear, but some testers actually showed anger and even happiness, as a sign of relief for performing the task correctly.

Finally, it was observed that, to obtain a good emotion detection accuracy, it was necessary to mount correctly the Meta Quest Pro headset, since incorrect mounting affects capturing of facial expressions.

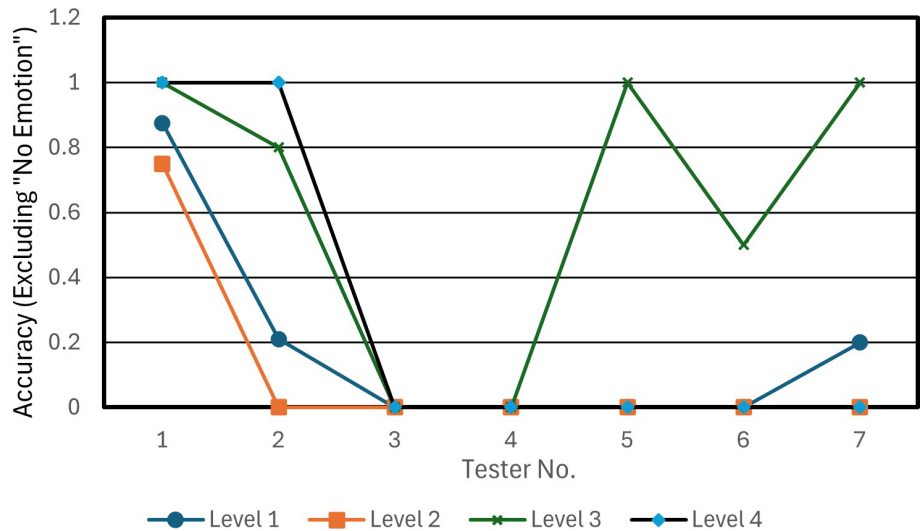

Figure 16: Emotion detection accuracy for each tester excluding "No Emotion" events.

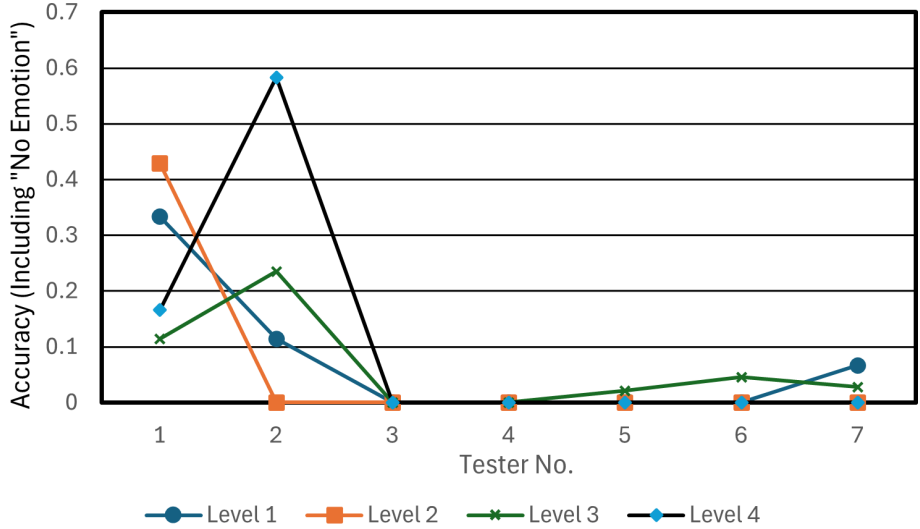

Figure 17: Emotion detection accuracy for each tester including "No Emotion" events.

*5.3.3. Emotions and user performance*

The emotional responses captured during the performed fire drill simulations were categorized into three groups: good emotions (e.g., happiness, contempt), bad emotions (e.g., anger, fear, disgust) and no emotions expressed at all. Thus, the obtained results allow for evaluating the overall experience of each tester for each of the four levels of the simulation.

Figure 18 shows the percentage of detected good emotions for each user and level. It can be observed that positive emotions are in general very low for most of the levels. This was expected, since a ship fire drill involves certain level of stress that cannot be directly linked to good feelings (except for situations where relief is expressed). In any case, there are some testers like 2 and 7 that in Figure 18 show a significantly high level of good emotions (33.33%) in specific levels. Tester 4 also showed a relatively high percentage of good emotions (up to 14.58%). The rest of the testers had little to no positive emotions throughout the whole simulation.

On the other hand, negative feelings were more prominent, as demonstrated in Figure 19. For instance, Tester 2 showed the highest level of negative emotional states, with 66.67% at level 2 and 58.33% at level 4. Tester

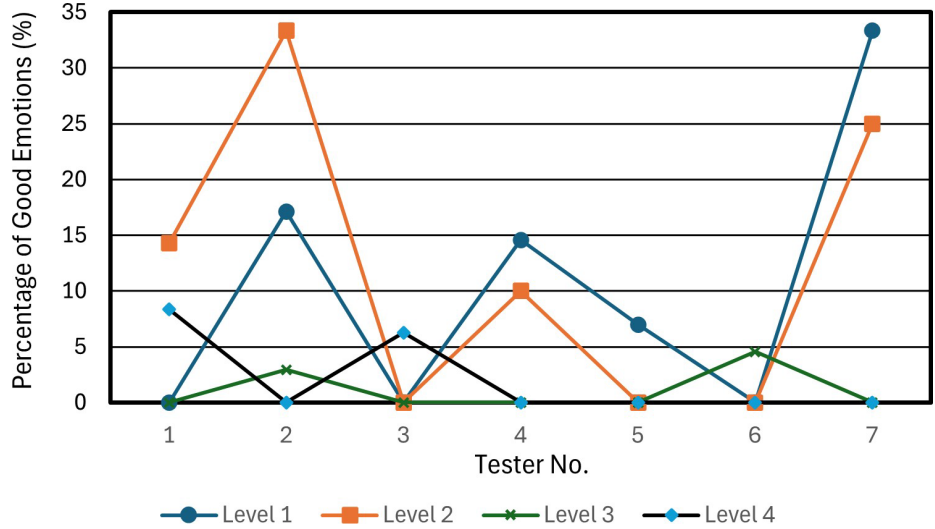

Figure 18: Percentage of good emotions experienced by each tester through each level.

1 was close behind, with 42.86% at level 2 and 38.10% at level 1. Testers 3 and 7 had only average levels of negative emotions, while testers 4, 5 and 6 displayed little to no negative emotional responses.

Finally, Figure 20 shows that a significant proportion of the testers, expressed high intensities of no emotions across all rating levels. For instance, Tester 3 exhibited a lack of emotion exceeding 90% across all levels, with the exception of level 4, where the percentage fell to 81.25%. Likewise, Testers 5 and 6 consistently recorded a 100% absence of emotion in multiple levels. In contrast, Testers 1 and 2 demonstrated diminished levels of emotional detachment as the simulation advanced, with Tester 1 decreasing from 61.90% in level 1 to 42.86% in level 2, suggesting a rise in emotional involvement.

Overall, the obtained results suggest that there was a clear division between the testers who felt emotional during the simulation and those who did not. Such a lack of emotional involvement can be related to multiple factors related to each individual, as well as to the fact that the testers considered the task more like a game rather than an actual fire drill, which might have affected their sense of presence (their awareness of being inside a virtual environment) [75].

Moreover, it is worth indicating that this article is focused more on de-scribing the design, development and demonstration of the viability of detecting emotions in a practical Metaverse application, so the size of the test group and their profile (e.g., naval engineers) condition the potential results. Thus, although it is out of the scope of this article, future research should be aimed at evaluating larger and more diverse testing groups.

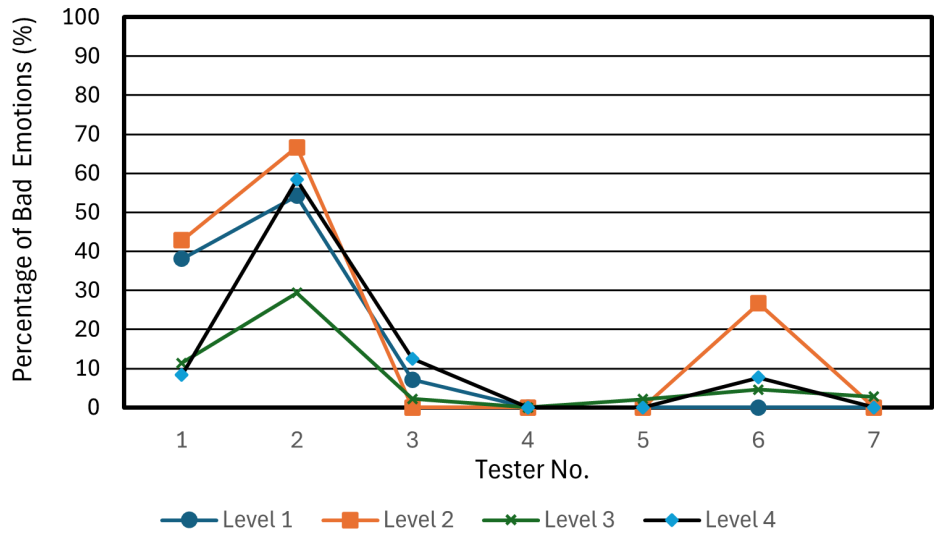

Figure 19: Percentage of bad emotions experienced by each tester through each level.

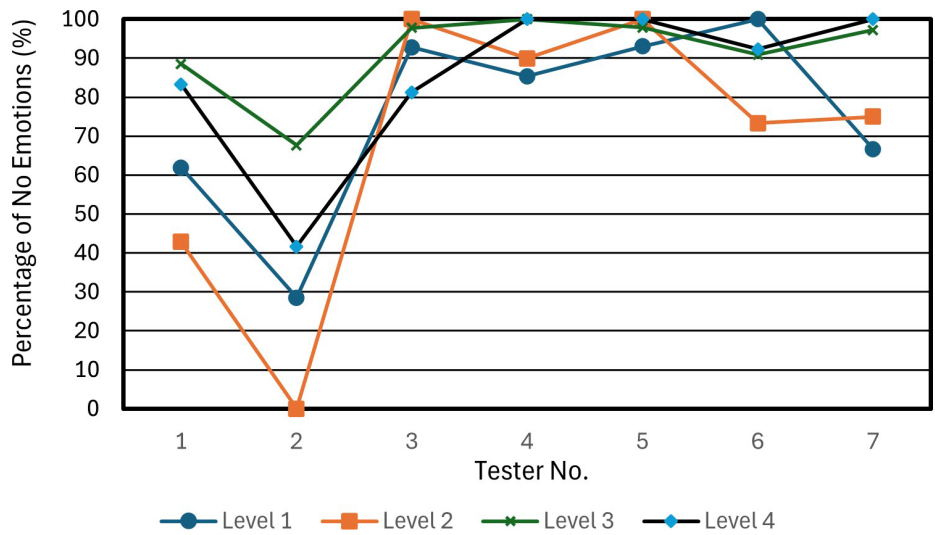

Figure 20: Percentage of "No Emotion" captured for each tester through each level.

## 6. Key findings

The path to the Metaverse still needs to be paved to provide truly immersive experiences that consider humans as their center, both in terms of user experience (i.e., interactions) and real response (i.e., reactions). After carrying out the development and evaluation of the ship fire drill application, the following key findings can be indicated to future researchers and Metaverse developers:

- It is currently possible to create Metaverse applications for training future Meta-Operators in critical tasks like disaster relief scenarios where fires occur.

- Although technologies like VR have progressed significantly in the last years and have become more affordable, they still need to be enhanced in terms of graphics to provide more realism in certain situations. Moreover, such technologies need to be enhanced to overcome a problem that is more concerning: during the two performed experiments, five testers experienced slight to severe motion sickness. This indicates that completely immersive technologies may not be adequate for certain people or they require some level of adaptation to use them.

- Emotional response is not easy to predict. An example of this fact is illustrated in Table 4, where there is a notable difference between the expected emotional responses and the ones actually detected when a tester had to found and extinguish a fire. Of the seven testers, only Tester 1 was surprised during all four levels of the test. Tester 2 went from an indifferent attitude into a fearful one in the engine room setting. The rest of the testers showed indifference throughout the simulation, which is opposite to the expected fear or surprise. This indicates that the simulator did not induce a strong emotional response on the testers, but, if that is a requirement for other applications, additional immersive artifacts should be included.

- In relation to the previous point, it is worth noting that every tester showed different responses in the same fire drill scenario, even when they were experiencing a similar situation. The results show that there were testers that showed no specific emotions, but it must be noted that such a response is conditioned by the scenario, the personality and background of each tester, the correct mounting of the headset and by how FACS define emotion detection. These factors allow for concluding that, although headsets like Meta Quest Pro enable detecting facial expressions, it is necessary to calibrate the system for every person before its actual training to detect emotions accurately. Moreover, external hardware (e.g., IoT sensors) can also be a good complement to facial expression detection.

- Although the performed tests allowed for demonstrating that a future emotion-aware Metaverse can be possible, the limited sample size impedes to extract deeper conclusions. For instance, such a reduced size influences the conclusions drawn from the second experiment, since the testers who re-took this simulation seemed less emotionally involved than those taking it for the first time. This could be interpreted as a bit of emotional desensitization, wherein continued access to the simulation may have diminished a perception of urgency or seriousness associated with a fire scenario. It is important to note that other factors like cultural and social norms also affect the emotional responses obtained by a test group [76].

- To obtain better performance results, future Metaverse application developers should consider that the higher the realism, the better the engagement of the users and the lesser the approach to the simulator as a form of entertainment. The prospects for realism can be enhanced by the development of multisensory feedback and personalization. Developers should also introduce variability to prevent desensitization and prioritize emotional realism through dynamic AI-driven interactions. By focusing on these areas, future virtual environments can foster stronger emotional connections and more impactful experiences.

## 7. Conclusions

This article introduced an XR fire drill simulator designed for maritime safety training that is able to monitor the interactions of the users, what they stare at and to detect their emotions. By integrating emotion detection technologies such as eye tracking and facial expression analysis, the system aimed to enhance realism, evaluate trainees' stress responses and improve decision-making in high-pressure fire emergencies.

| Tester No. | Level 1     | Level 2     | Level 3     | Level 4     | Expected Emotion |
|------------|-------------|-------------|-------------|-------------|------------------|
| 1          | Surprise    | Surprise    | Surprise    | Surprise    | Surprise/Fear    |
| 2          | Indifferent | Indifferent | Fear        | Fear        | Surprise/Fear    |
| 3          | Indifferent | Indifferent | Indifferent | Indifferent | Surprise/Fear    |
| 4          | Indifferent | Indifferent | Indifferent | Indifferent | Surprise/Fear    |
| 5          | Indifferent | Indifferent | Indifferent | Indifferent | Surprise/Fear    |
| 6          | Indifferent | Indifferent | Indifferent | Indifferent | Surprise/Fear    |
| 7          | Indifferent | Indifferent | Indifferent | Indifferent | Surprise/Fear    |

Table 4: Testers' emotional responses when locating and extinguishing the fire across different levels.

After providing a detailed review on the state of the art and background work, the article described the main components of the developed emotion-aware XR application. The performed experiments allowed for measuring the amount of time spent in carrying out certain tasks, concluding that trainees with prior VR or gaming experience navigated the environment more efficiently. Moreover, the introduction of task-tracking visuals and navigation guidance significantly improved user performance, reducing task completion times between 14.18% and 32.72%, thus enhancing procedural adherence. Emotional response varied widely among participants: while some exhibited engagement, others remained indifferent, likely treating the experience as a game rather than a critical safety drill. In addition, the carried out emotion-based performance analysis indicates that fear and surprise were not consistently elicited as expected, suggesting a need for more immersive elements, such as haptic feedback, auditory cues or environmental changes that simulate real-world urgency.

In conclusion, this article provided, through the description of a fire drill simulator, detailed guidelines for future researchers and developers on how to create the next generation of emotion-aware Metaverse applications.

**Funding**

This work has been funded by grant TED2021-129433A-C22 (HELENE) funded by MCIN/AEI/10.13039/501100011033 and the European Union NextGenerationEU/PRTR.

**Ethics statement**

This study does not involve human participants, animal experiments, or any other procedures requiring ethical approval.